\documentclass[11pt]{article}
\usepackage{geometry}                % See geometry.pdf to learn the layout options. There are lots.
\usepackage{graphicx}
\usepackage{amssymb}
\usepackage{epstopdf}
\usepackage{amsmath, amssymb, graphics}
\newcommand{\mathsym}[1]{{}}

\title{The role of Spectator Fragments at an electron Ion collider.}
\author{Sebastian White\\
Physics Department, Brookhaven National Lab, Upton, NY 11973\\
Mark Strikman\\
Physics Department, Pennsylvania State University, University Park, PA 16802}
\date{}                                           % Activate to display a given date or no date

\begin{document}
\maketitle

\begin{abstract}
%%%I think most of people write EIC not eIC
Efficient detection of spectator fragments is key to the main topics at an electron-ion collider (eIC). Any process which leads to
emission of fragments or $\gamma$'s  breaks coherence in diffractive processes. Therefore this is equivalent to
non-detection of rapidity gaps in pp collisions. For example, in coherent  photoproduction of vector mesons their 
4-momentum transfer distribution would image the "gluon charge" in the nucleus in the same way that Hofstadter
measured its charge structure using elastic scattering of $\sim$100 MeV electrons. Whereas he could measure the $\sim$4 MeV 
energy loss by the electron due to excitation of nuclear energy levels (Figure 1), even the energy spread of the incident beam
would prevent such an inclusive selection of quasielastic events at an eIC.
	The only available tool is fragment detection. Since, in our example, one finds that $\sim100\%$ of deexcitations go through
$\gamma$'s or 1 neutron, rarely to 2 neutron and never to protons(due to Coulomb barrier suppression), the eIC
design should emphasize their detection.
Also in incoherent diffraction fragmentation involves emission of a neutron.
%We discuss the role of Ion beam fragments at electron ion colliders with an emphasis on the implications for design of the experiments.
%We discuss experience from Heavy Ion collisions and lower energy phenomena.
\end{abstract}

\section{Introduction}

	A future electron-ion collider would significantly advance our understanding of nuclear structure and QCD. Designs for such accelerators are already far advanced so it is possible to say things about how to incorporate sensitivity to fragments from the breakup of nuclei in such collisions.
	
	In this paper we discuss two benchmark physical processes that are likely to be the focus of an electron ion program- quasielastic vector meson production and deeply inelastic  jet production. 
	%(we do not discuss an interesting physics related to the fragmentation of protons  in $ep$ and deuterons in $e^2H$ collisions). 
	
	In Figures 2,3 we show both incoming ion and electron beam interacting with a nuclear target to emphasize the fact that at electron ion colliders we face very similar
challenges to the ones at heavy ion colliders (see eg. \cite{SVW}). For this reason the lessons from heavy ions are directly relevant to our problem.

\begin{figure}
\centering
\includegraphics[width=0.5\linewidth]{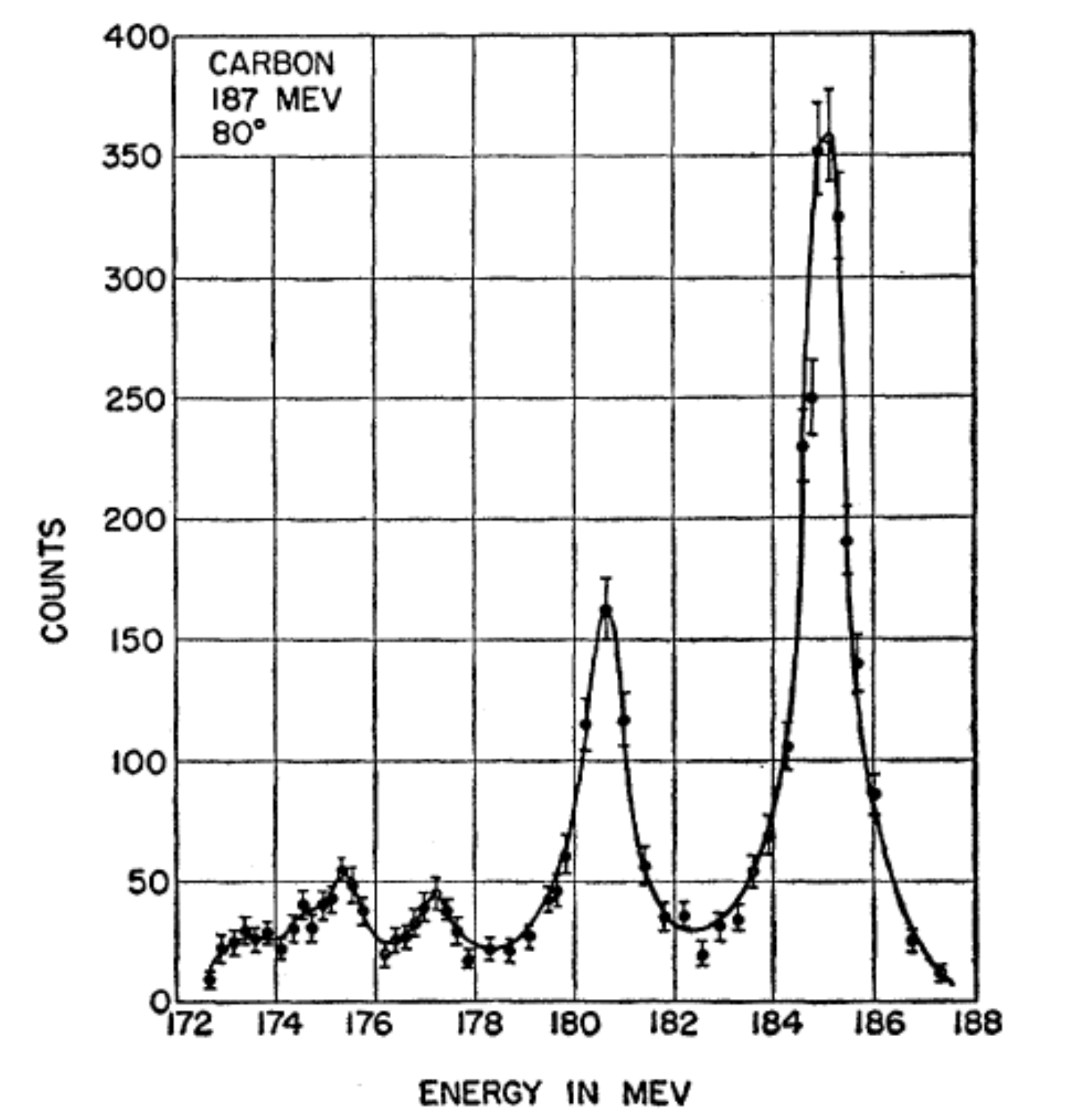}
\caption{Hofstadter's data were used to measure the charge profile of the carbon nucleus. The background from inelastic 
scattering was eliminated by removing events with $E_{beam}-E_{electron}\geq$ 1.5 MeV.}
\label{fig:carboninel}
\end{figure}

\subsection{Vector Meson Production}
	
	In the first process $e + A \to e + VM + A$, depicted in Fig.2, a small color dipole exchanges 
	%a pair of gluons 
	a two gluon ladder 
	(with quantum numbers of the vacuum) which interact coherently with the nucleus. At sufficiently large $Q^2$ (or for sufficiently large mass of the quarks of the onium)  the t-distribution of the vector mesons measures directly the gluon nuclear profile in a way analogous to the measurement of the electromagnetic distribution in Rutherford scattering. Breakup of the nucleus destroys coherence. Analogously to Bjorken's rapidity gap method to resolve coherent processes in proton collisions, detection of nuclear fragments has been used in the measurement of vector meson photoproduction in ("Ultraperipheral") heavy ion collisions\cite{PHENIXUPC}.
	
	There are two distinct types of incoherent processes which are characterized by different challenges for fragmentation measurements. We discuss them below.

%electro is addded since for EIC it is a more generic process - for UPC it would be photoproduction only

\subsection{Hard electro/photoproduction}
	
	The second process, depicted in Fig. 3, is deeply inelastic scattering off the nucleus, which will probe the partonic structure of nuclei. At HERA two aspects of this process were measured- inclusive and diffractive structure functions, which probe different topics in QCD. In the second case %process 
	the target nucleus  remains intact so that one could think of the process as a transition of the virtual photon to  hadronic configurations which 
	%exchanges a colorless object which interacts with the incoming nucleus.
	 %In these diffractive interactions the "Pomeron"
	 %%better to avoid Pomeron here since even in the soft physics language it is not a single Pomeron but multipomeron exchange 
	 interact coherently with the entire nucleus. Again the most powerful tag of coherence in this process is the absence of beam fragments in the final state.
\begin{figure}
\centering
\includegraphics[width=0.5\linewidth]{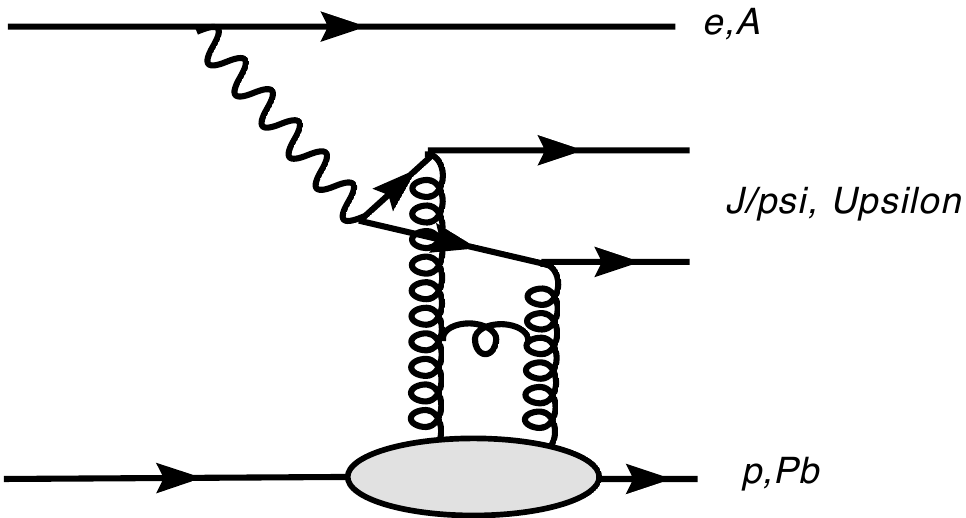}
\caption{Diffractive Vector Meson production.}
\label{fig:vecmeson}
\end{figure}

\begin{figure}
\centering
\includegraphics[width=0.5\linewidth]{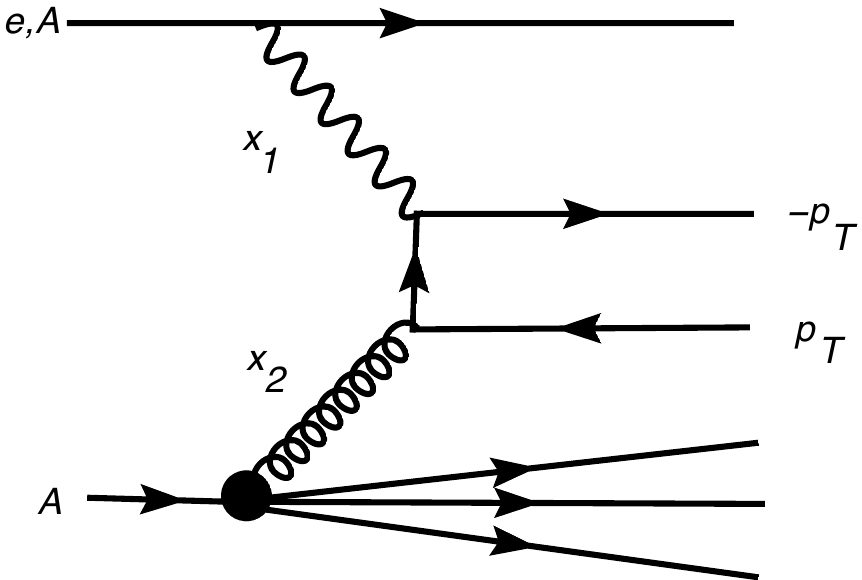}
\caption{Hard jet photoproduction.}
\label{fig:dis}
\end{figure}
\section{Heavy Ion Fragmentation}

	A great deal about ion beam fragmentation has been learned from the  past 30 years of Heavy Ion Physics. For our purposes the most significant features are;
\begin{itemize}
\item{the types of fragments- particularly their charge to mass ratio (Z/A) since this determines where they could be detected.}
\item{the multiplicity of each species}
\item{their momentum distributions}
\end{itemize}

	For each of these there are differences from the dominant characteristics of hard nuclear collisions, which generally result in a larger number of fragments.
	The composition of fragments in nucleus-nucleus collisions was measured by NA49\cite{NA49}. They find a neutron to proton ratio which ranges from 1.3 to 1.9 over a calculated range of impact parameters from 2.1 to 8.4 fermi. At small impact parameters a small fraction of the forward energy is carried by fragments with $A>1$. Although these are expected to be primarily deuterons or tritons the composition wasn't directly measured. For our purposes even fragments with Z/A=1/2 would be difficult to detect, as they were in NA49, because they would travel a short distance in the beampipe close to the beam before colliding with the beam walls.
		
	The composition from fragments in eA collisions will differ from the NA49 one depending on the process of interest. The difference originates from a much smaller number of wounded nucleons and from the ability of the virtual photons to penetrate deep into the nucleus. Also, the composition of fragments may depend on x, since at small x the virtual photon is likely to interact with several nucleons, making a significant fraction of fragmentation processes resemble pion  - nucleus interactions 
	
	The momentum distribution of nuclear fragments consists of several components\cite{STRIKWHITE}. In hard nuclear collisions the fragment nucleons have approximately a simple Fermi distribution. %Detailed m
	Measurements in which fragmentation of light nuclei (oxygen) was studied
favored a Gaussian momentum distribution( Feschbach-Huang\cite{Feshbach}) with the same rms as the Fermi one.
	However available Heavy Ion data cover a limited range in fragment $p_T$ whereas data from the Brookhaven AGS\cite{Tang:2002ww,Piasetzky:2006ai}  and JLab \cite{Subedi:2008zz,Egiyan:2005hs}   have demonstrated that short range correlations result in a significant hard component in the $p_T$ spectrum. On the other hand particle emission on the outer side of the nuclear fragment is expected to be significantly different from these spectator characteristics. There particle emission will resemble the evaporation spectrum seen in low energy collisions 
	
	We are incorporating a comprehensive description of the fragmentation distributions in the HIJING event generator\cite{HIJING}.
	
\subsection{electron Ion fragmentation spectra}

	In process 1) there are several contributions 
	to the cross section which one would like to distinguish from the coherent process.
	
	(a) Excitation of nuclear levels.  This process is known to be important both for eA elastic scattering and for pA scattering for medium and heavy nuclei. In the case of Pb there are numerous transitions to  excited levels with energies below the break up threshold of about 8 MeV. As a result, though these contributions %not 
constitute a $\le 10\%$ correction to the total elastic cross section, they  exceed the elastic cross section starting at $t$ corresponding to the first diffractive minimum of the elastic cross section.	To suppress this contribution one has to measure photons with energies of a few MeV in the nucleus rest frame (100- 400 MeV for nuclear beams with E/A=100GeV). They are emitted with a mean angle of 10 mrad. in the lab frame. In the case of Uranium beams the excitation threshold is much smaller $\sim 1 MeV$, hence this background  may be smaller (see discussion below).
	
	(b) Inelastic processes of incoherent scattering off individual nucleons  with elastic or quasielastic nucleon final state. In this case
	fragmentation following a low energy nuclear excitation and also reinteractions of low energy nucleons in the nucleus  will be predominantly 
	%strictly
	 via evaporation and, because of the nuclear coulomb barrier, the proton component is strongly suppressed- ie $\frac{n}{p} \sim$10. 
	 This type of inelastic diffraction is interesting in itself as it is sensitive to the dynamics of the small dipole nucleus interaction. The process is $\propto A$ for the regime of color transparency, and with the approach to the black disk regime the A-dependence should become weaker reaching ultimately $A^{1/3}$ - scattering off the rim of the nucleus.
	% The second type of inelastic process is incoherent photoproduction which is interesting in itself because.......... In these events a
%there is coherent exchange leaving an intact nucleon which recoils and heats up the nucleus. Here too the primary spectrum is an evaporation one with $\sim 6$MeV kinetic energy. The scattered nucleon could also be detected and it will have a $p_T$-distribution with roughly $\sim Exp{-6p_T}$.

The spectrum of neutrons in these events should be predominantly 
 evaporation  with an average kinetic energy of about 6 MeV. In a small fraction of events ( $\propto  A^{-1/3}$)  when the process occurred near the nuclear surface, the  recoil nucleon will be able to 
 escape and be produced with the same t spectrum as in the elementary reaction.
 
 %The scattered nucleon could also be detected and it will have a $p_T$-distribution with roughly $\sim Exp{-6p_T}$.

It is worth noting also that separation of diffractive and inclusive DIS events for $x\le 0.01$ is of crucial importance for probing the onset of the regime of strong absorption in the DIS interaction in which most of  the $q\bar q$ and $q\bar q$g components of the virtual photon 
are interacting with the nucleus with the strength close to the black disc regime. In such a regime coherent diffraction would constitute half of the cross section (In the current models the actual number for the top eRHIC energies  and $Q^2\sim4GeV^2$ is 20$\div$ 35 \%.
To do a measurement of the difffraction/total cross section ratio a measurement of rapidity gaps and neutron emission would be crucial. (The excitation of the levels is
a small correction in this case as one is interested in the integral of the cross section over t.)
\section{Lessons from Ion Colliders}

	At RHIC spectator measurements have played a central role in collisions involving all species from deuterons to Gold. Specifically, in hard collisions, spectator neutrons  (measured with Zero Degree Calorimeters(ZDC)) were used primarily to determine centrality and the reaction plane. In peripheral heavy ion collisions PHENIX measured the process shown in Fig. 2 and used the measurement of neutron multiplicity to detect rapidity gaps in the events\cite{PHENIXUPC}. 
	%%since you procedure is different from STAR's for rho's better to add few words.
	Currently PHENIX is analyzing a larger sample where it should be possible to use the neutron spectators, measured in the ZDC, to extract the incoherent cross section. PHENIX also measured the electromagnetic breakup cross section in deuteron-Au collisions detecting the neutrons from the deuteron\cite{Glauber}. 
	
	Protons from beam fragmentation in collisions with nuclei are not captured in the beamline as illustrated in figure 4. The bending power of the insertion dipole is much larger for protons(1) than for heavy nuclei (ie Z/A(Au)=0.4 , Z/A(Deuteron)=0.5). It might seem that a proton detector could be inserted between this dipole and the start of the beamline. However RF impedence considerations would make this difficult.
	
	It might seem that an interaction region designed without an insertion dipole would allow one to measure protons in the beamline. However there would be many
problems with such a design. Firstly the protons are not, in any case, captured because the momentum dispersion function, $\eta(s)$  rapidly reaches a value of $>$0.5m 
in both RHIC and MeRHIC( a lower energy stage with a 4 GeV electron beam)
%%MeRHIC  not defined
 optics so
protons would be lost in any practical accelerator. Secondly one would lose sensitivity to all neutrals and deuteron-like fragments. Some versions of the MeRHIC design
all but eliminate this insertion dipole and we recommend strongly against that. The RHIC collisions geometry, instead, is a good one.
	
	PHENIX installed a forward proton calorimeter to detect proton fragments. This calorimeter was used for the analysis of deuteron photodissociation where proton energy resolution of $20\%$ was obtained as can be seen in figure 5. Since there would be some interest in proton detection also at an eIC it would be interesting to evaluate
potential calorimeter technologies since it would be easy to do better than $20\%$ for 100 GeV protons.

	Position sensitivity of the calorimeter is also interesting, since in the vertical direction it measures $p_T$ and in the horizontal plane a combination of $p_T$ and
momentum loss
%%%probably better  to write p_t^{(y)} or simply p_y
\begin{equation}
		\delta y=\frac{p_T^y }{100GeV}\times L_{effective}
\end{equation}
\begin{equation}
		\delta x=\frac{p_T^x }{100GeV}\times L_{effective}+\eta(s)\times\frac{\delta p}{100}
\end{equation}
\begin{figure}
\centering
\includegraphics[width=0.5\linewidth]{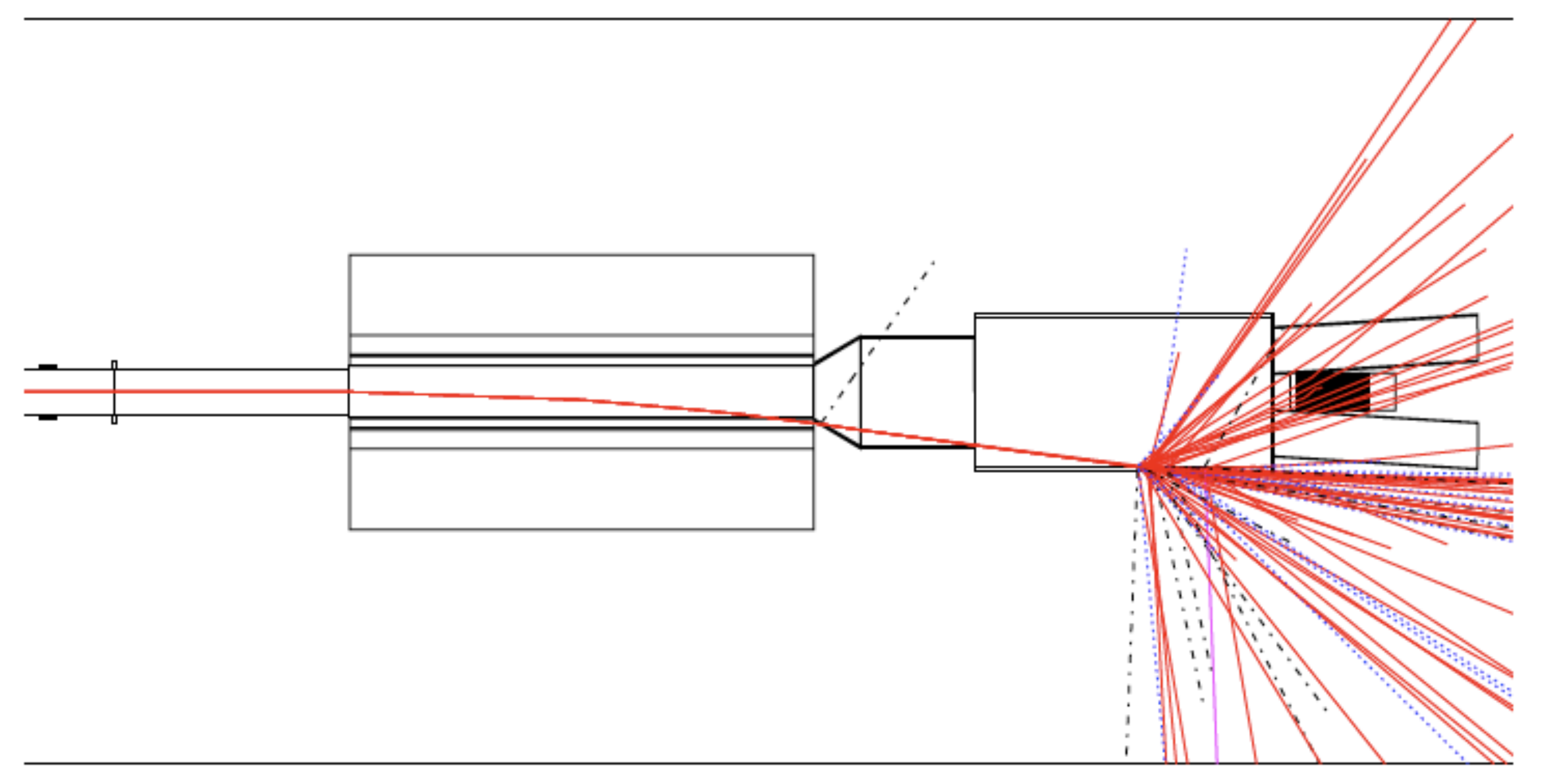}
\caption{Trajectory of protons at an ion collider.}
\label{fig:pwall}
\end{figure}
\begin{figure}
\centering
\includegraphics[width=0.5\linewidth]{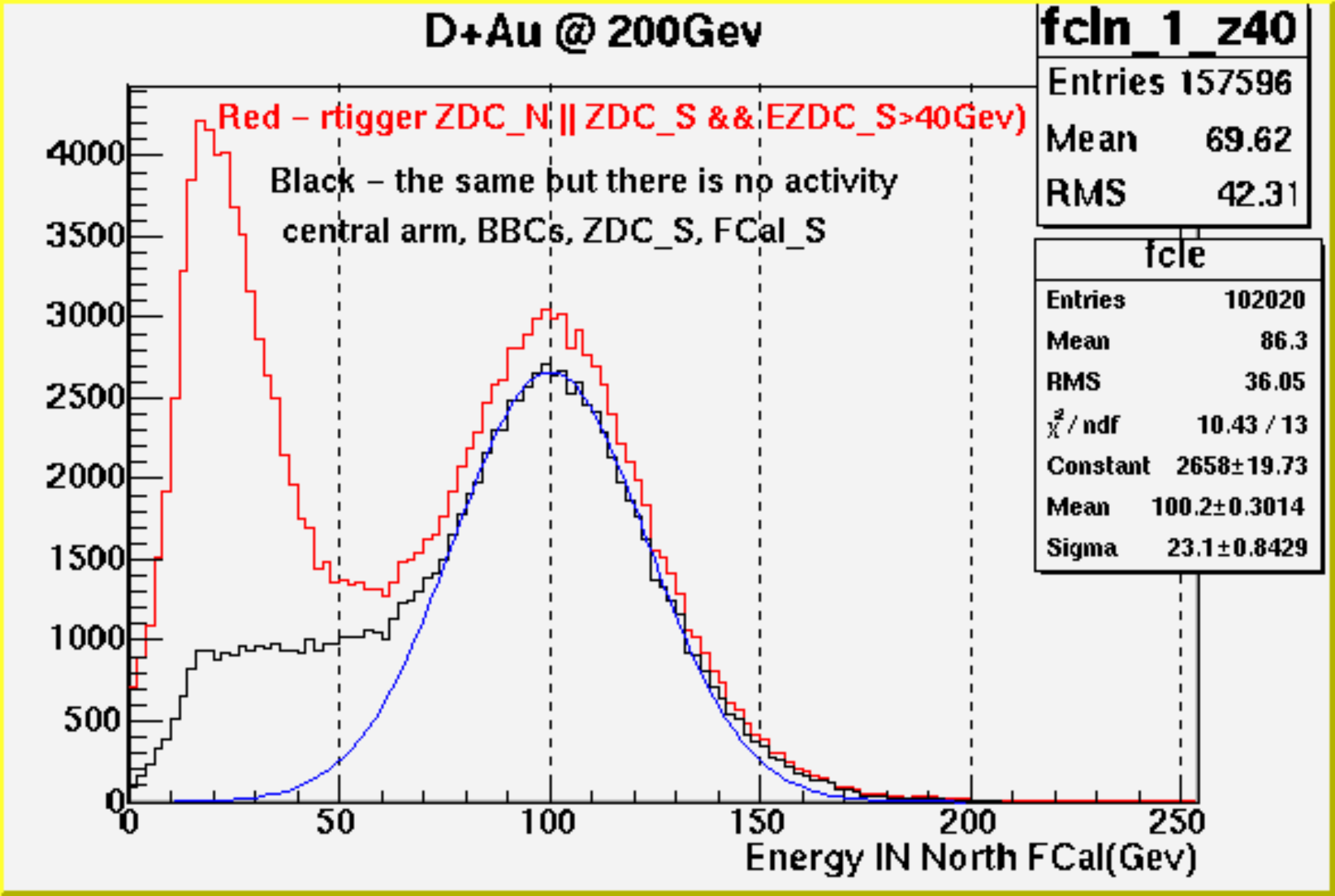}
\caption{Performance of PHENIX forward proton calorimeter in the deuteron photodissociation analysis\cite{Glauber}. The predominant
background is the deuteron stripping process in which the proton breaks up. Therefore there's a low energy peak.}
\label{fig:fcaln}
\end{figure}	
\section{Possible advantages in %Mark's idea 
using Uranium beams }
As we already mentioned, measurement of coherent scattering off heavy nuclei at $-t =p_t^2 \ge 0.02 GeV^2$ that is near and beyond the first minimum, presents a challenge due to the contribution of the excited levels. At the same time
in the case of $^{238}U$ the situation may be more promising because the threshold for fission is close to 1  MeV which means a typical energy transfer $E=-t/2m_N$ for $-t\ge 0.02 GeV^2$
 \footnote{We thank M.Zhalov for the suggestion that it maybe advantageous to use uranium beams to suppress background due to the excitation of the nuclear levels.}  Fission leads to the production of two nuclear fragments with 
$A \sim 100$ and $\sim$ 3 neutrons. One needs to perform a 
theoretical
 study to check whether 
a low fission threshold is sufficient to suppress the contribution of level excitations below 1 MeV, however this appears pretty likely. 

\begin{figure}
\centering
\includegraphics[width=0.5\linewidth]{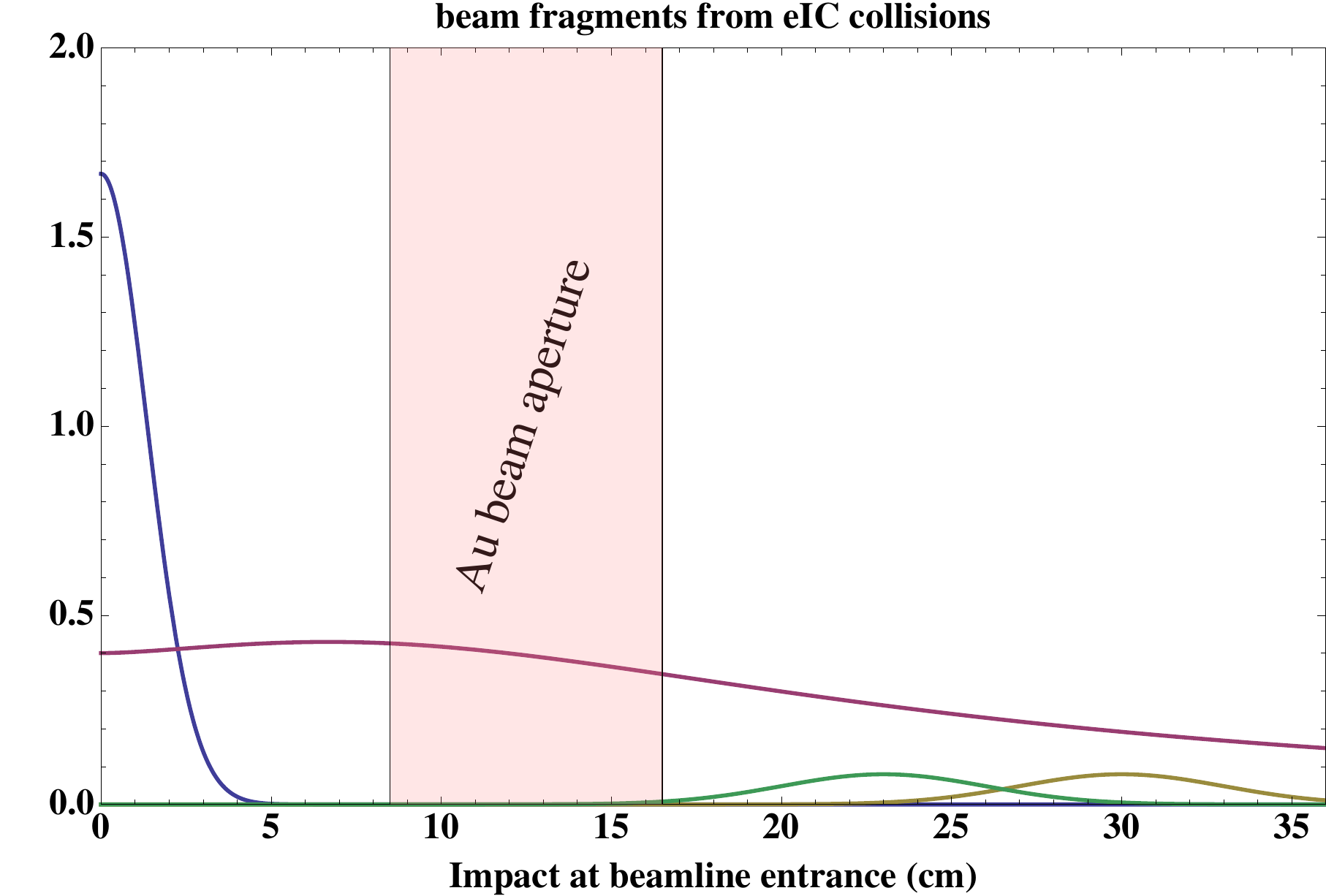}
\caption{Distribution of fragments at the entrance to the beamline in the RHIC geometry. The size of the beamline aperture is indicated. Neutrons (blue) and photons (red) are centered on the forward direction and completely dominate the fragmentation. Charged fragments, deuterons (green) and protons (yellow) are rarely produced due to Coulomb barrier supression. In a low energy eIC design the bending power of the DS magnet is lower by a factor of 4. Then all fragments enter the beamline. Charged fragments, if produced, would immediately be lost to collisions with the beampipe since the momentum dispersion, $\eta$(s) is larger than 0.5 meters.}
\label{fig:pwall}
\end{figure}

\section{Fragmentation of light nuclei}
Reactions with production of $J/\psi$ could be employed to study the gluon field in light nuclei like $^4He, ^2H$. In this case there are no excited states below threshold and one can either try to detect the nucleus or look to putting a veto on the breakup of the deuteron to pn or the break up of $^4He$ which is obviously more challenging for channels like $^3H p, ^2H^2H$. At the same time one can try to switch to the study of the break up channel say for $^2H$ 

\begin{equation}
e^2H \to e J/\psi + pn
\end{equation} 

The study of quasielastic scattering on a nucleon with large momentum belonging to a short-range correlation in the deuteron (or in a light nucleus like $^3He, ^3He$) will allow one to check whether the transverse gluon distribution in the bound nucleon deviates from that in a free nucleon. It is expected that such effects should increase approximately as the square of the spectator nucleon momentum in the nucleus rest frame.   This will be analogous for the gluon channel to the recent observations of the deviation of the $G_{Ep}/G_{Mp}$ form factor ratio from the free one in the $e^4He \to e p^3H$ reaction studied at Jlab \cite{Jlab}.

Such a measurement will require good resolution (say better than 10\% in the energy of the proton and neutron) which would allow one to check exclusivity of the process.

One can also find a kinematic domain where this process
will measure the simultaneous interactions of a small dipole with two nucleons which is expressed at high energies through a non-diagonal triple Pomeron interaction.  

\section{Connection to the needs of ep mode}
One of the novel directions of EIC in the ep mode is the study of phenomena involving fragmentation of protons. A sample of such 
 processes are fragmentation of polarized protons in $\vec{e}\vec{p} \to p (n, \Lambda) + X$, long range correlations between the production of hadrons in the target and current fragmentation region which is sensitive to the transverse momentum correlations in the proton. Studies of proton fragmentation were performed at HERA for the unpolarized case. A wide range of $x_F$ down to  $x_F\sim 0.3 $ was covered. However the specifics of HERA kinematics did not allow them to explore the dependence of the proton fragmentation on $x$ which is expected to lead to a more steep drop of the spectrum at large $x_F$ with increasing x \cite{Frankfurt:1981mk}
.

Good acceptance in this kinematics is also necessary for the study of nondiagonal GPDs, both in exclusive reactions like $ep\to e \pi^- \Delta^{++}$,
 see the review in \cite{Goeke:2001tz}
and in the branching 2$\to $ 3 processes \cite{Kumano:2009he}, like
$\gamma p \to \pi^+ \pi^0 +n, \gamma p \to \pi^+ \pi^+ + \Delta^-$, etc.


\begin{thebibliography}{10}

\bibitem{SVW}
M. ~Strikman, R. ~Vogt, and S. ~White, Phys. Rev. Lett. 96, 082001 (2006).
arXiv:hep- ph/0508296
\bibitem{E814}
J.~Barrette  {\it et al.},
	BNL-44931.
\bibitem{Evaporation}
M. ~Strikman, M.~G. ~Tverskoy , M.~B.~Zhalov,
arxiv.org/abs/nucl-th/9806099.
\bibitem{NA49}
H. ~AppelshAauser {\it et al.},
Eur. Phys. J. A 2, 383 (1998).
\bibitem{STRIKWHITE}
M.~Strikman, S.~White, PHENIX Internal Note.
arXiv:hep- ph/0508296
\bibitem{Frankfurt:1981mk}
  L.~L.~Frankfurt and M.~I.~Strikman,
  Phys.\ Rept.\  {\bf 76}, 215 (1981).
\bibitem{Frankfurt:2008zv}
  L.~Frankfurt, M.~Sargsian and M.~Strikman,
  %``Recent observation of short range nucleon correlations in nuclei and their
  %implications for the structure of nuclei and neutron stars,''
  Int.\ J.\ Mod.\ Phys.\  A {\bf 23}, 2991 (2008)
  [arXiv:0806.4412 [nucl-th]].
  \bibitem{PHENIXUPC}
  PHENIX Collaboration,  Phys. Lett. B 679, 321 (2009).
  \bibitem{Glauber}
  S.~White, "Diffraction Dissociation- 50 Years Later", DIS05 Proceedings.
  arXiv:nucl-ex/0507023v2
  \bibitem{Tang:2002ww}
  A.~Tang {\it et al.},
  %``n p short-range correlations from (p,2p + n) measurements,''
  Phys.\ Rev.\ Lett.\  {\bf 90}, 042301 (2003)
  [arXiv:nucl-ex/0206003].
\bibitem{Piasetzky:2006ai}
  E.~Piasetzky, M.~Sargsian, L.~Frankfurt, M.~Strikman and J.~W.~Watson,
  %``Evidence for the Strong Dominance of Proton-Neutron Correlations in
  %Nuclei,''
  Phys.\ Rev.\ Lett.\  {\bf 97}, 162504 (2006)
  [arXiv:nucl-th/0604012].
\bibitem{Subedi:2008zz}
  R.~Subedi {\it et al.},
  %``Probing Cold Dense Nuclear Matter,''
  Science {\bf 320}, 1476 (2008)
  [arXiv:0908.1514 [nucl-ex]].
\bibitem{Egiyan:2005hs}
  K.~S.~Egiyan {\it et al.}  [CLAS Collaboration],
  %``Measurement of 2- and 3-Nucleon Short Range Correlation Probabilities in
  %Nuclei,''
  Phys.\ Rev.\ Lett.\  {\bf 96}, 082501 (2006)
  [arXiv:nucl-ex/0508026].
  \bibitem{Feshbach}
H.~ Feshbach and K.~ Huang,  
Phys. Letters 47B (1973) 300.
\bibitem{Goldhaber}
A.~Goldhaber,
	Physics Letters 53B (1974) 306.

\bibitem{Jlab}M.~Paolone, S.~P.~Malace, S.~Strauch and f.~t.~E.~Collaboration,
  %``Polarization Transfer in the 4He(e,e'p)3H Reaction at Q^2 = 0.8 and 1.3
  %(GeV/c)^2,''
  arXiv:1002.2188 [nucl-ex].
\bibitem{Frankfurt:1981mk}
  L.~L.~Frankfurt and M.~I.~Strikman,
  %``High-Energy Phenomena, Short Range Nuclear Structure And QCD,''
  Phys.\ Rept.\  {\bf 76}, 215 (1981).
\bibitem{Goeke:2001tz}
  K.~Goeke, M.~V.~Polyakov and M.~Vanderhaeghen,
  %``Hard Exclusive Reactions and the Structure of Hadrons,''
  Prog.\ Part.\ Nucl.\ Phys.\  {\bf 47}, 401 (2001)
  [arXiv:hep-ph/0106012].

\bibitem{Kumano:2009he}
  S.~Kumano, M.~Strikman and K.~Sudoh,
  %``Novel two-to-three hard hadronic processes and possible studies of
  %generalized parton distributions at hadron facilities,''
  Phys.\ Rev.\  D {\bf 80}, 074003 (2009)
  [arXiv:0905.1453 [hep-ph]].
  \bibitem{HIJING}
 M.~Alvioli, M.~Vargyas,T.~Csorgo,A.~Ster, M.~Strikman and S.~White,
  %``Novel two-to-three hard hadronic processes and possible studies of
  %generalized parton distributions at hadron facilities,''
 paper in preparation.
\end{thebibliography}
\end{document}